\documentclass[twocolumn,twoside,slac_two]{revtex4}
\usepackage{graphicx}
\usepackage{fancyhdr}
\usepackage{hyperref}
\hypersetup{
    colorlinks=true,
    urlcolor=magenta,
}
\def \aap  {A\&A}

\def \aj  {AJ}

\def \aplett  {ApJL}

\def \jcap  {JCAP}
\def \prd {Phy. Rev. D}

\def \mnras {MNRAS}
\def \nat {Nature}
\def \physrep {Phys. Rept.}

\begin{document}

\title{Anisotropies in GeV-TeV cosmic ray electrons and positrons}

%

\author{S. Manconi}
\affiliation{INFN Torino, via Pietro Giuria 1, 10125 Torino, Italy}
\author{M. Di Mauro}
\affiliation{ W. W. Hansen Experimental Physics Laboratory, Kavli Institute for Particle Astrophysics and Cosmology, Department of Physics and SLAC National Accelerator Laboratory, Stanford University, Stanford, CA 94305, USA}

\author{F. Donato}
\affiliation{INFN Torino, via Pietro Giuria 1, 10125 Torino, Italy}
\begin{abstract}
High energy cosmic ray electrons and positrons probe the local properties of 
our Galaxy. In fact, electromagnetic energy losses limit the typical 
propagation scale of GeV-TeV electrons and positrons to a few kpc. In the 
diffusion model,  nearby and dominant sources may  produce an observable dipole 
anisotropy in the cosmic ray  fluxes. 
We present a detailed study on the role of anisotropies from nearby sources in 
the interpretation of the observed GeV-TeV cosmic ray electron and positrons 
fluxes. 
We compute predictions  for  the anisotropies from known astrophysical sources 
as supernova remnants and pulsar wind nebulae of the ATNF catalog. 
Our results are compared with current anisotropy upper limits from the Fermi-
LAT, AMS-02 and PAMELA experiments.
\end{abstract}

\maketitle

\thispagestyle{fancy}


\section{Introduction} 

In the last years, high precision measurements of the fluxes of electrons and positrons ($e^\pm$) in Cosmic Rays (CRs) have been performed 
 by the AMS-02 \cite{PhysRevLett.110.141102,PhysRevLett.113.121101}, \textit{Fermi}-LAT \cite{2012PhRvL.108a1103A} and PAMELA \cite{2009Natur.458..607A} experiments. 
 The observed fluxes can be interpreted as the emission from a variety of astrophysical sources in the Galaxy. 
In addition, the detected fluxes have been recently analyzed in terms of anisotropies in their arrival directions. Searches for anisotropies in the electron plus positron ($e^- + e^+$) flux \cite{2010PhRvD..82i2003A}, the positron to electron ratio \cite{PhysRevLett.110.141102} and the positron ($e^+$) flux \cite{Adriani:2015kfa} have been presented respectively from the \textit{Fermi}-LAT, AMS-02 and PAMELA experiments, all ending up with upper limits on the dipole anisotropy. 
In this work we discuss how the search for anisotropies in the $e^\pm$  fluxes at GeV-TeV energies can be an interesting tool, in addition to the measured fluxes, to study the properties of near sources, as for example near SNRs. 

\section{The model: sources and propagation in the Galaxy} \label{sec:modeling} 
The production of CR $e^\pm$ in our galaxy is possible through different processes.
\textit{Primary} $e^-$ are accelerated with Fermi non relativistic shocks up to high energies in SNRs \cite{1987PhR...154....1B}. 
In addition, both $e^+$ and $e^-$ are produced in the strong magnetic fields that surround pulsars and then accelerated with relativistic shocks in the PWN \cite{2011ASSP...21..624B}.
A source of \textit{secondary} $e^\pm$ is  the fragmentation of primary CR nuclei in the interstellar medium material. 
Indipendent of the production mechanism, $e^\pm$ propagate in the Galaxy and are affected by a number of processes. 
Above a few GeV, the propagation is dominated by the diffusion in the interstellar magnetic field irregularities and by energy losses, which are due to inverse Compton scattering on ambient photons and to synchroton emission. This is tipically described by a diffusion equation for the number density $\psi = \psi(E, \mathbf{x}, t)\equiv dn/dE$ per unit volume and energy:
\begin{equation}
 \frac{\partial \psi}{\partial t}  - \mathbf{\nabla} \cdot \left\lbrace K(E)  \mathbf{\nabla} \psi \right\rbrace + 
 \frac{\partial }{\partial E} \left\lbrace \frac{dE}{dt} \psi \right\rbrace = Q(E, \mathbf{x}, t)
 \label{eq:diff}
\end{equation}
where $K(E)= K_0 E^\delta$ is the  diffusion coefficient, $dE/dt$  accounts for the energy losses and  $Q(E, \mathbf{x}, t)$ includes all the possible sources. 
In this work a semi analytical approach is followed to solve the diffusion equation in Eq.~\ref{eq:diff} for each source, as fully detailed in  ~\cite{2010A&A...524A..51D,new}. 
Within this approach, the Galaxy is modeled as a cylinder of radius $r_{\rm disc}=20$~kpc and half height L.  
The parameters $K_0, \delta, L$ are usually constrained from boron over carbon ratio~(B/C). In particular, in the following we show results for the MAX benchmark model derived in  ~\cite{Donato:2003xg}.
At high energies ($E> 10 $~GeV) the $e^\pm$ that we observe are probes of the local Galaxy. In fact, for leptons the energy loss timescale is smaller than the diffusion timescale. As an example, for GeV-TeV $e^\pm$ and MAX propagation model, the propagation scale $\lambda$ is less than $\sim 5$~kpc. 
The interpretation of high energy $e^\pm$ is thus connected to the inspection of local sources. 
The chosen modeling of $e^\pm$ sources is functional to this aim, and is based on  ~\cite{new, 2010A&A...524A..51D, 2014JCAP...04..006D}. 
We include both single SNRs and PWNe, whose characteristics are taken directly from the existing catalog, and a distribution of SNRs,  described by average characteristics.
The secondary component  is taken from \cite{2014JCAP...04..006D}. 
The spatial distribution of SNRs is modeled with a smooth distribution of sources active beyond a radius $R_{\rm cut}$ from the Earth (\textit{far} SNRs), and  following the radial profile derived in \cite{2015MNRAS.454.1517G}. Instead, single sources taken directly from the Green Catalog \cite{Green:2014cea} are considered for $R\leq R_{\rm cut}$ (\textit{near} SNR). 
To inspect the role of single near SNRs we consider $R_{\rm cut}=0, 0.7$~kpc.
The PWNe component is computed taking  from the ATNF catalog \cite{2005AJ....129.1993M} the sources with ages $50$~kyr$<t_{\rm obs}<10000$~kyr, since the release of the accelerated $e^\pm$ pairs is estimated to start at least after $40-50$~kyr after the pulsar birth \cite{2011ASSP...21..624B}.  
The energy injection spectrum $Q(E)$ for both SNR and PWN is  $Q(E)= Q_{0} \left( \frac{E}{E_0}\right)^{- \gamma} \exp \left(-\frac{E}{E_c} \right) $
where $E_c=5$~TeV is the cutoff energy and $E_0=1$~GeV. The index of the energy spectrum is expected to be different for particles accelerated in SNRs ($\sim 2-2.5$) and PWNe ($<2$). 
The normalization of this spectrum is constrained using catalog quantities for single SNRs and PWNe, while using average population characteristics for the smooth SNR component.  
For a single PWN the normalization is obtained supposing that a fraction $\eta$ of the spindown energy of the pulsar $W_0$  is emitted in form of $e^\pm$ pairs.  The normalization for a single SNR is constrained with the radio flux, the distance and magnetic field of the remnant (see Eq.~50 in \cite{2010A&A...524A..51D}). 
As for the smooth SNR distribution, the normalization can be connected to average Galactic characteristics, as for example the mean energy released in $e^-$ per century $E_{\rm tot, SNR}$. 
\section{Anisotropy}
CR $e^\pm$ with  observed energies in GeV-TeV range originated from relatively nearby locations in the Galaxy. This means that it could be possible that such high energy $e^\pm$ originate from a highly anisotropic collection of nearby sources. Under this hypothesis, even after the diffusive propagation in the Galactic magnetic field is taken into account, a residual small dipole anisotropy should be present in the observed fluxes. 
In the assumption of one or few nearby sources dominating the CR flux at Earth, the dipole anisotropy is usually defined as 
\begin{equation}\label{eq:dip}
 \Delta = \frac{I_{max}- I_{min}}{I_{max}+ I_{min}}
\end{equation}
being $I_{max}$ ($I_{min}$) the maximum (minimum) CR intensity values. 
In a diffusive propagation regime this can be computed as  $\Delta= \frac{3 K}{c} \left| \frac{\nabla \psi}{\psi} \right| $ (see  ~\cite{1964Ginzburg}), where $K$ is the diffusion coefficient and $\psi$ is the solution to Eq.~\ref{eq:diff}. 
Moreover, if a collection of electron and/or positron sources is present, the intensity of the CR flux as a function of direction in the sky $\mathbf{n}$ is (see  ~\cite{1971ApL.....9..169S})
\begin{equation}\label{eq:intensity}
 I(E, \mathbf{n}) = \frac{c}{4\pi}\sum_i  \psi_i(E)(1+ \Delta_i\, \mathbf{n}\cdot \mathbf{r}_i/r_i)
\end{equation}
where the index $i$ runs over all the sources at position $\mathbf{r}_i$ with electron and/or positron number density $\psi_i(E)$, $\Delta_i = \frac{3 K(E)}{c} \frac{|\nabla \psi_i(E) |}{\psi_i(E)}$, and the total dipole anisotropy is computed directly by means of Eq.~\ref{eq:dip}. 
To compare our prediction with the present upper limits we compute the integrated dipole anisotropies as a function of the minimum energy $E_{\rm min }$. We integrate fluxes in Eq.~\ref{eq:dip} up to $5$~TeV. 
%
%
\section{Results}\label{sec:results} 
\begin{figure*}
\begin{center}
\includegraphics[width=0.425\textwidth]{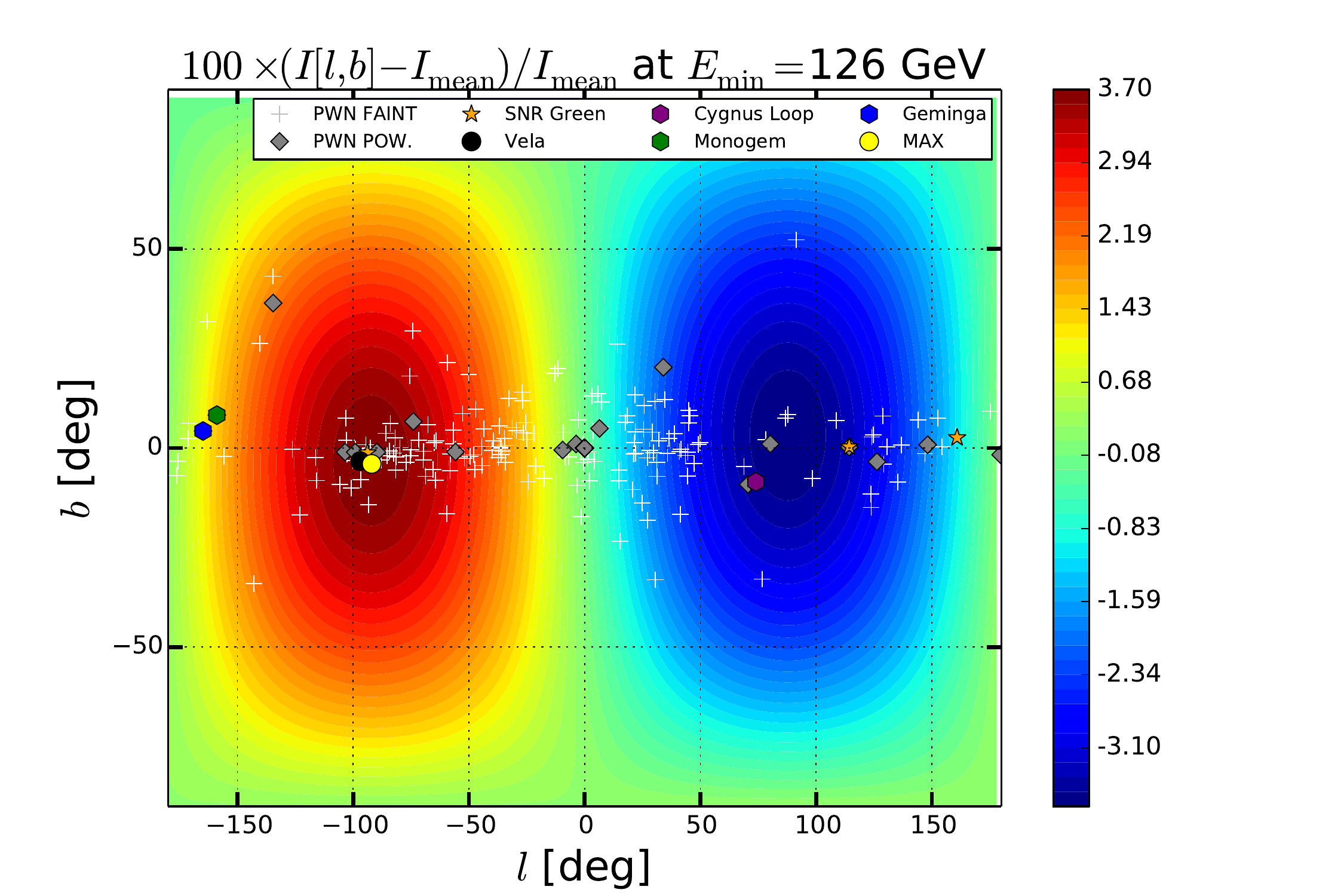}
 \includegraphics[width=0.425\textwidth]{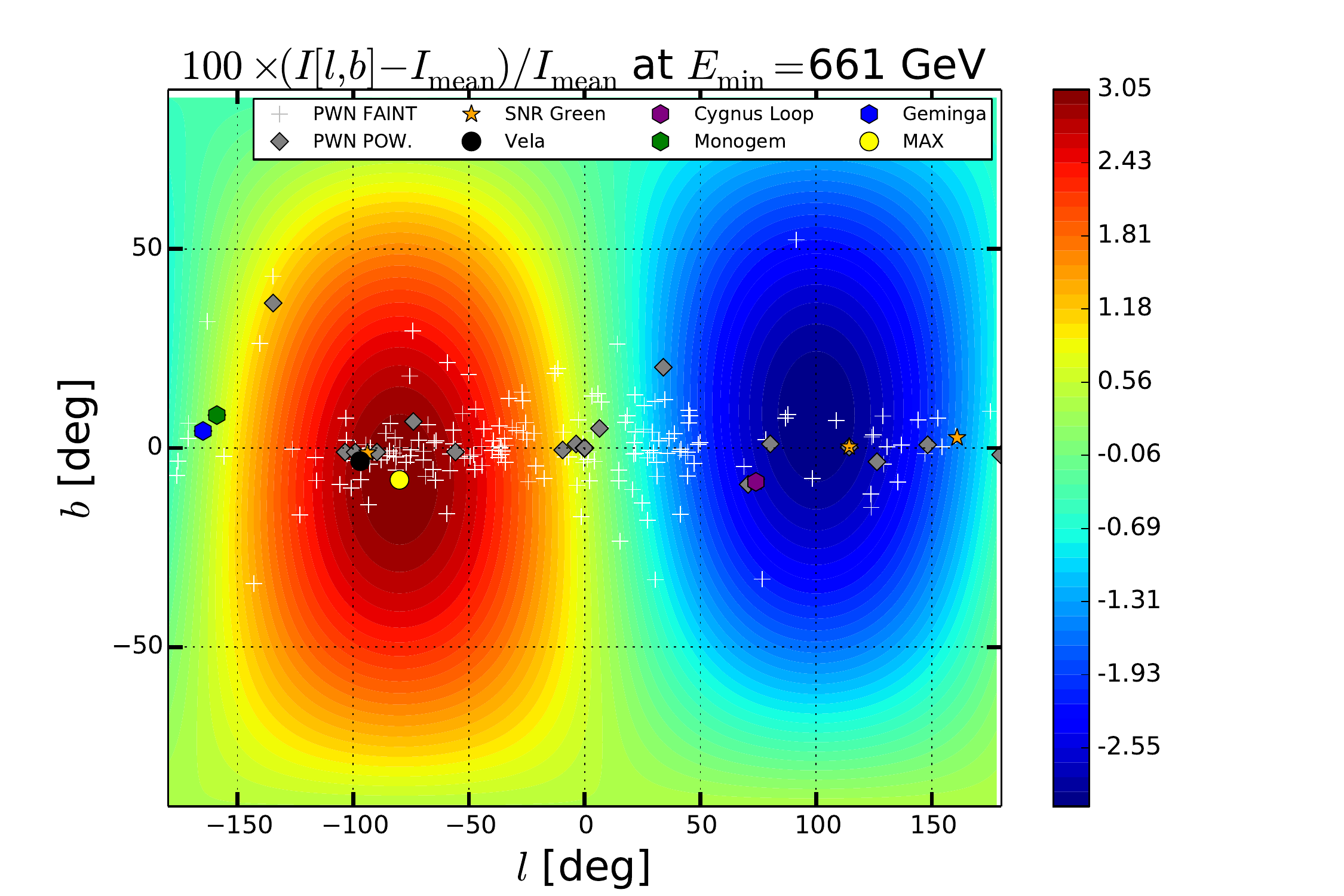}
 \includegraphics[width=0.425\textwidth]{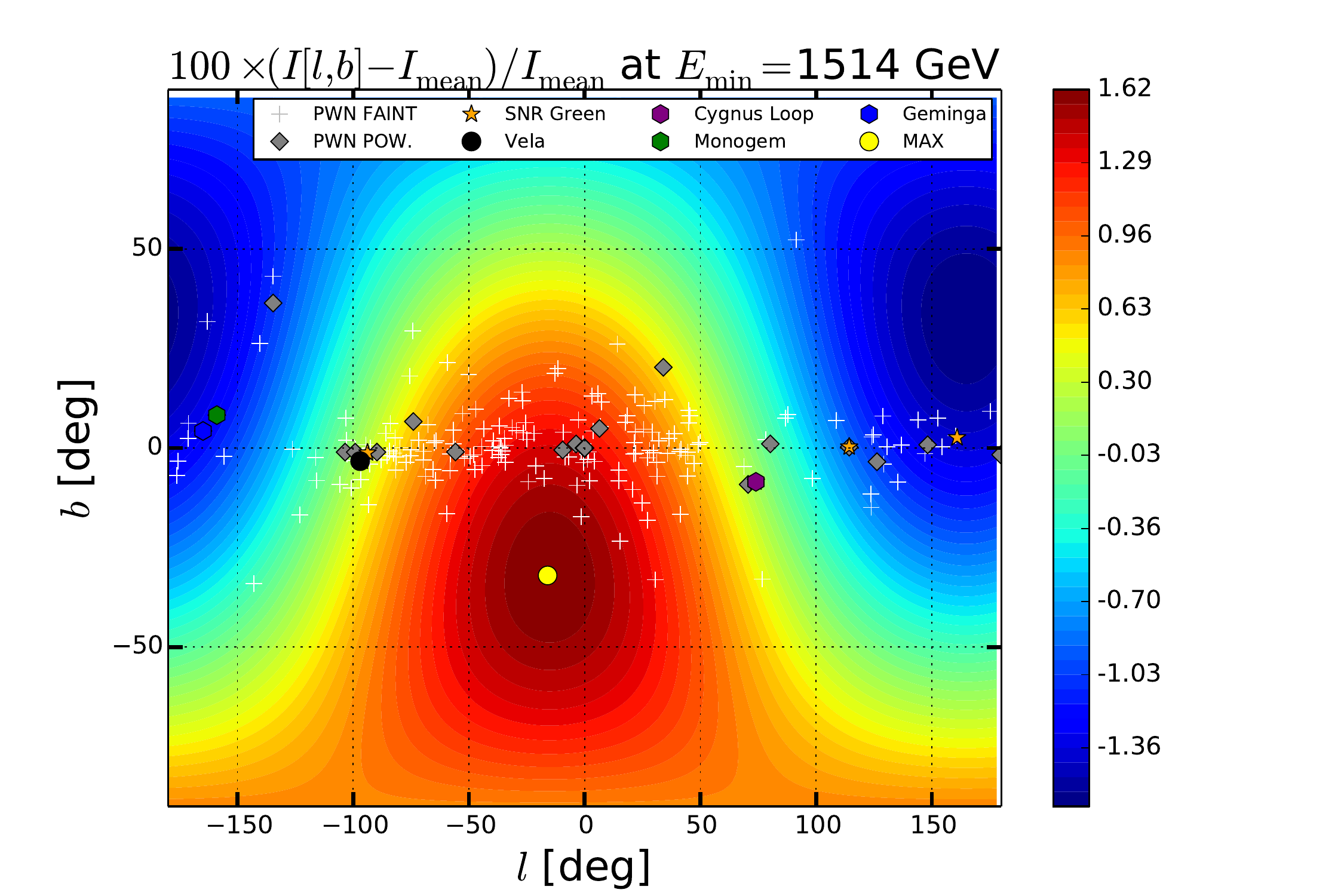}
\includegraphics[width=0.425\textwidth]{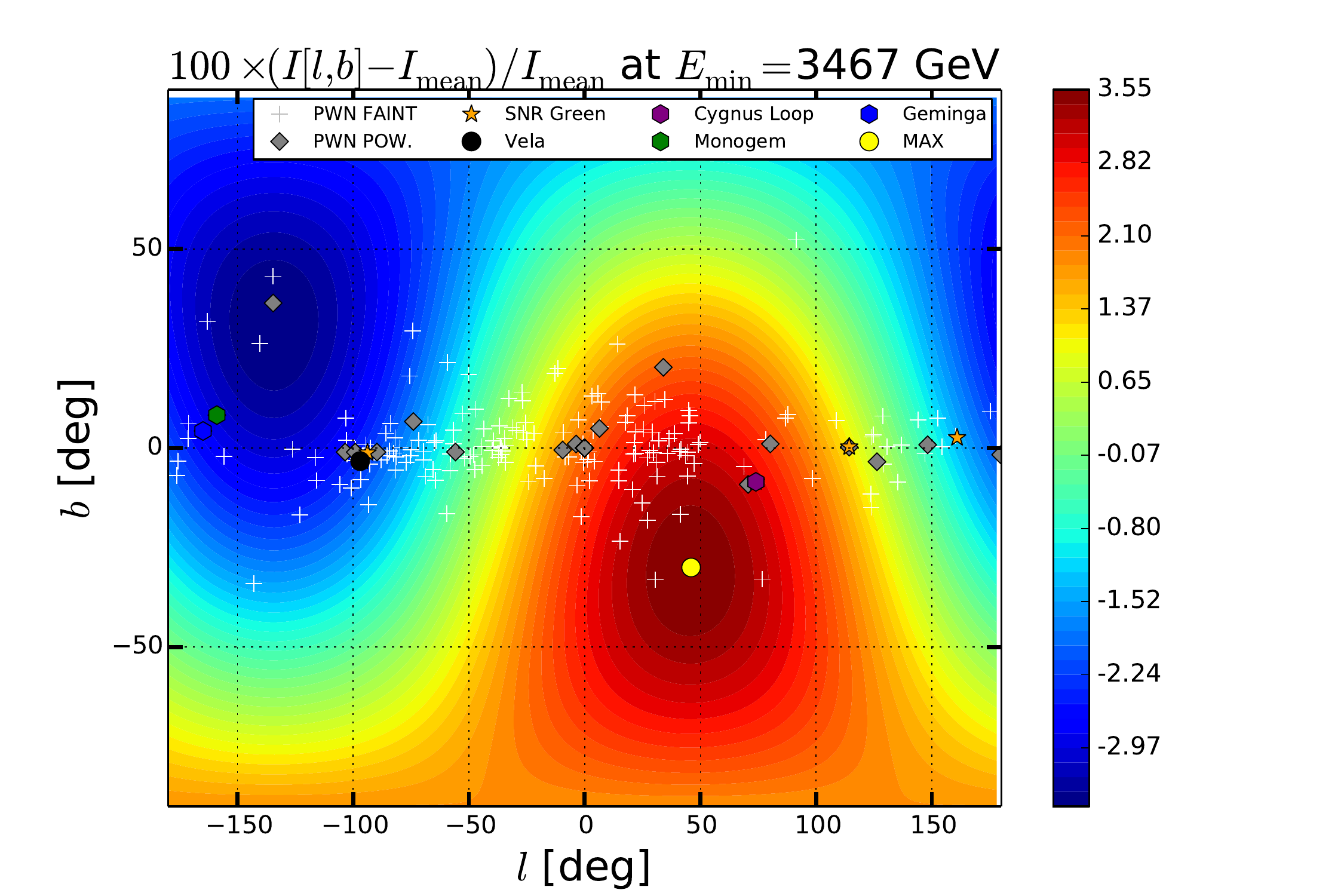}

\end{center}
\caption{\label{fig:countl} Countor plots of the intensity of the $e^+ + e^-$ flux as a function of the direction in the sky, obtained for all the sources in the model with $R_{\rm cut}=0.7$~kpc. 
The color scale indicates the percentage relative difference between the intensity in a given direction of the sky (l, b) [deg] and the mean intensity computed from the entire source collection.
The position of the maximum intensity is highlighted with a yellow dot, while the other symbols indicate the position of the sources.}
\end{figure*}

The aim of this analysis is to provide dipole anisotropies predictions for models compatible with the observed $e^\pm$ fluxes. 
Therefore, each model is fitted to the AMS-02 data on the $e^+$ \cite{2014PhRvL.113l1102A} and $e^+ + e^-$ \cite{2014PhRvL.113v1102A} fluxes.
Data are fitted starting from $E=10$~GeV. This choice minimizes the effect of the solar modulation of fluxes, that is however taken into account with a modulation potential $\phi_F$, according to the force field approximation.
The inspected models differ mainly for the treatment of the contribution from local sources.
As an example, two models and the corresponding predictions for anisotropies are discussed here. 
\begin{figure}
\begin{center}
\includegraphics[width=0.425\textwidth]{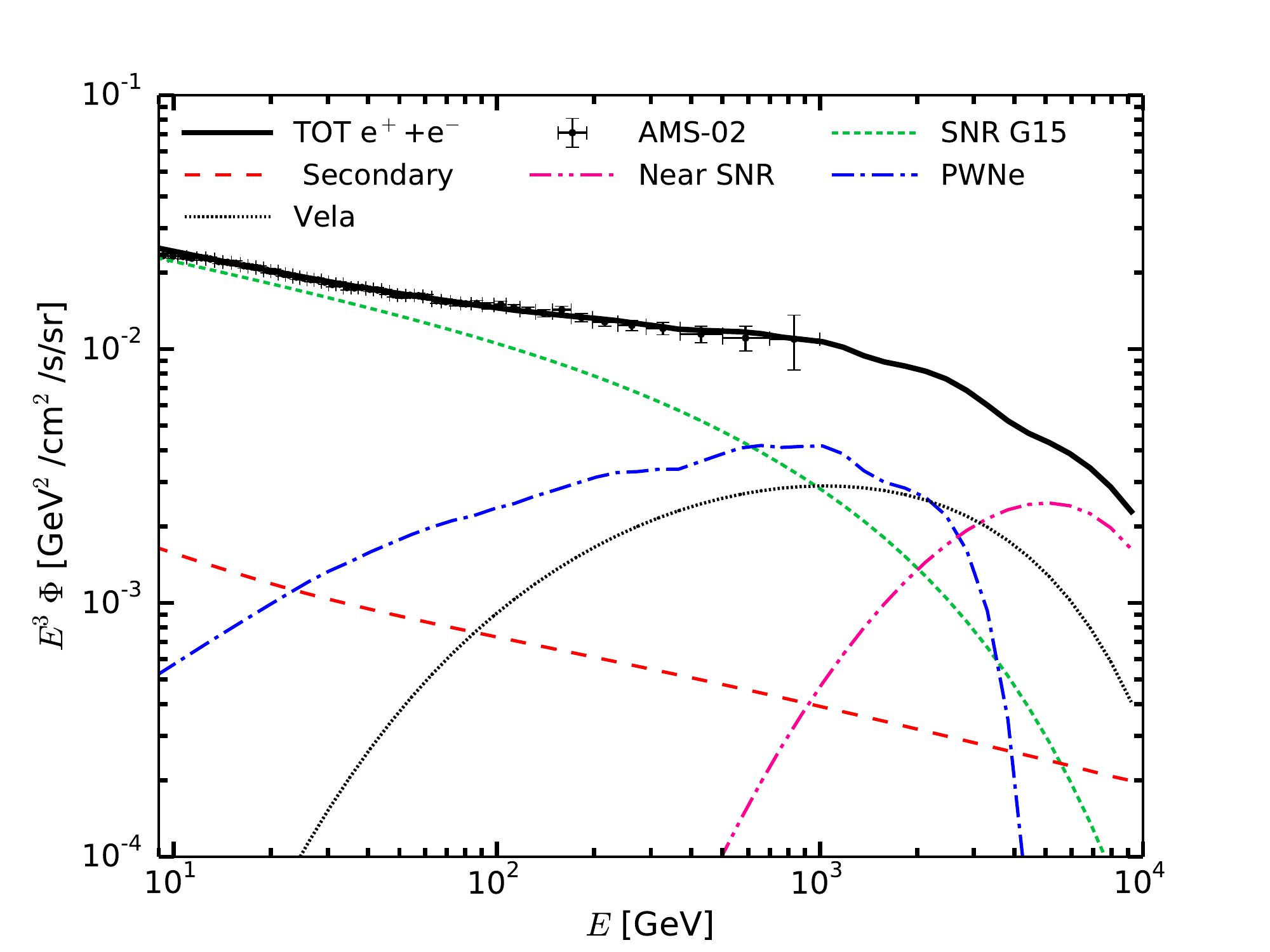}
\includegraphics[width=0.425\textwidth]{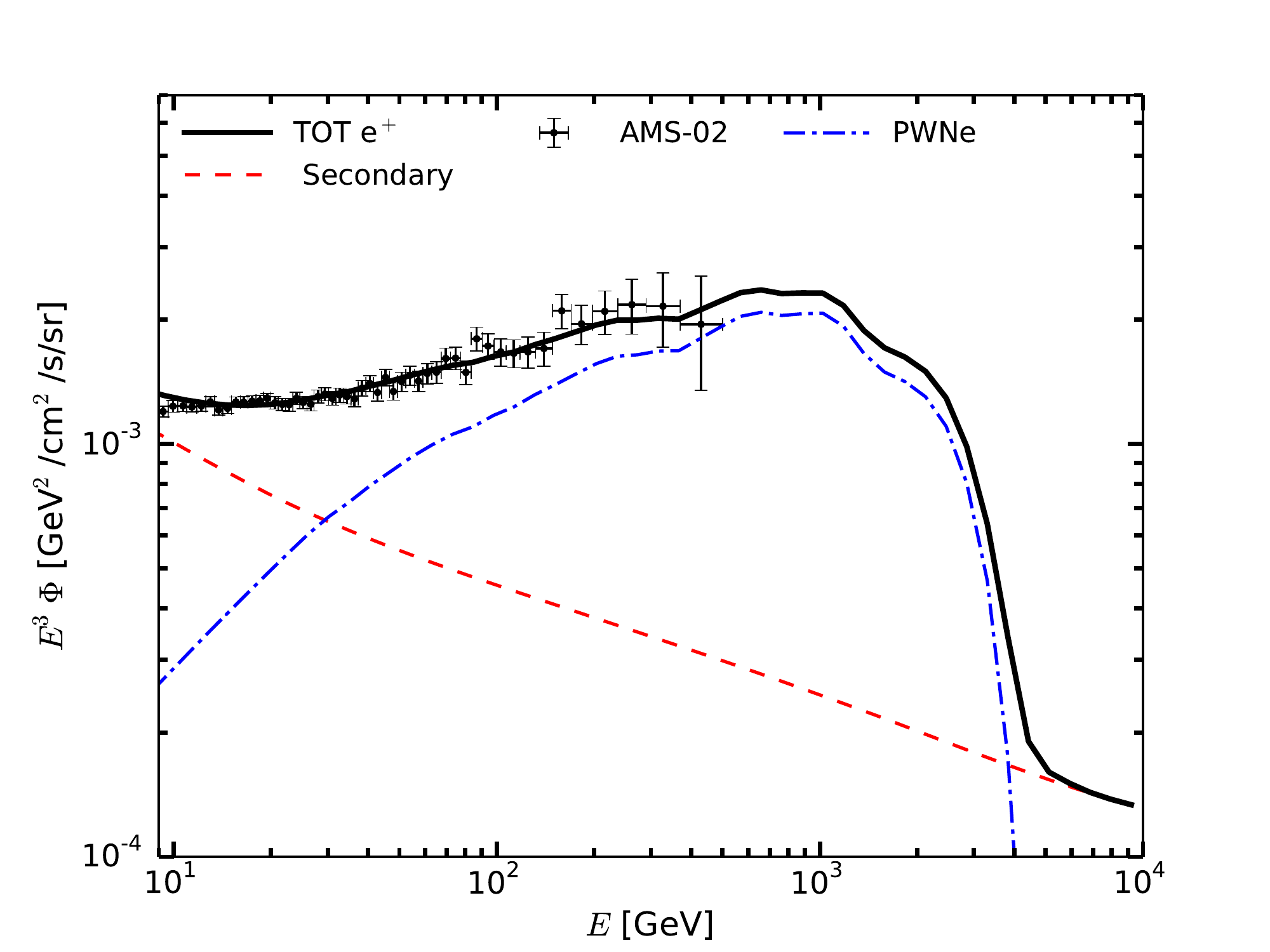}
\end{center}
\caption{\label{fig:fluxes} 
Best fit to $e^+ + e^-$ (upper panel) and $e^+$ (lower panel) AMS-02 data \cite{2014PhRvL.113l1102A, 2014PhRvL.113v1102A} for the model with $R_{\rm cut}=0.7$~kpc described in Sec. \ref{sec:results} and MAX propagation setup. 
}
\end{figure}
\begin{figure}
\begin{center}
\includegraphics[width=0.425\textwidth]{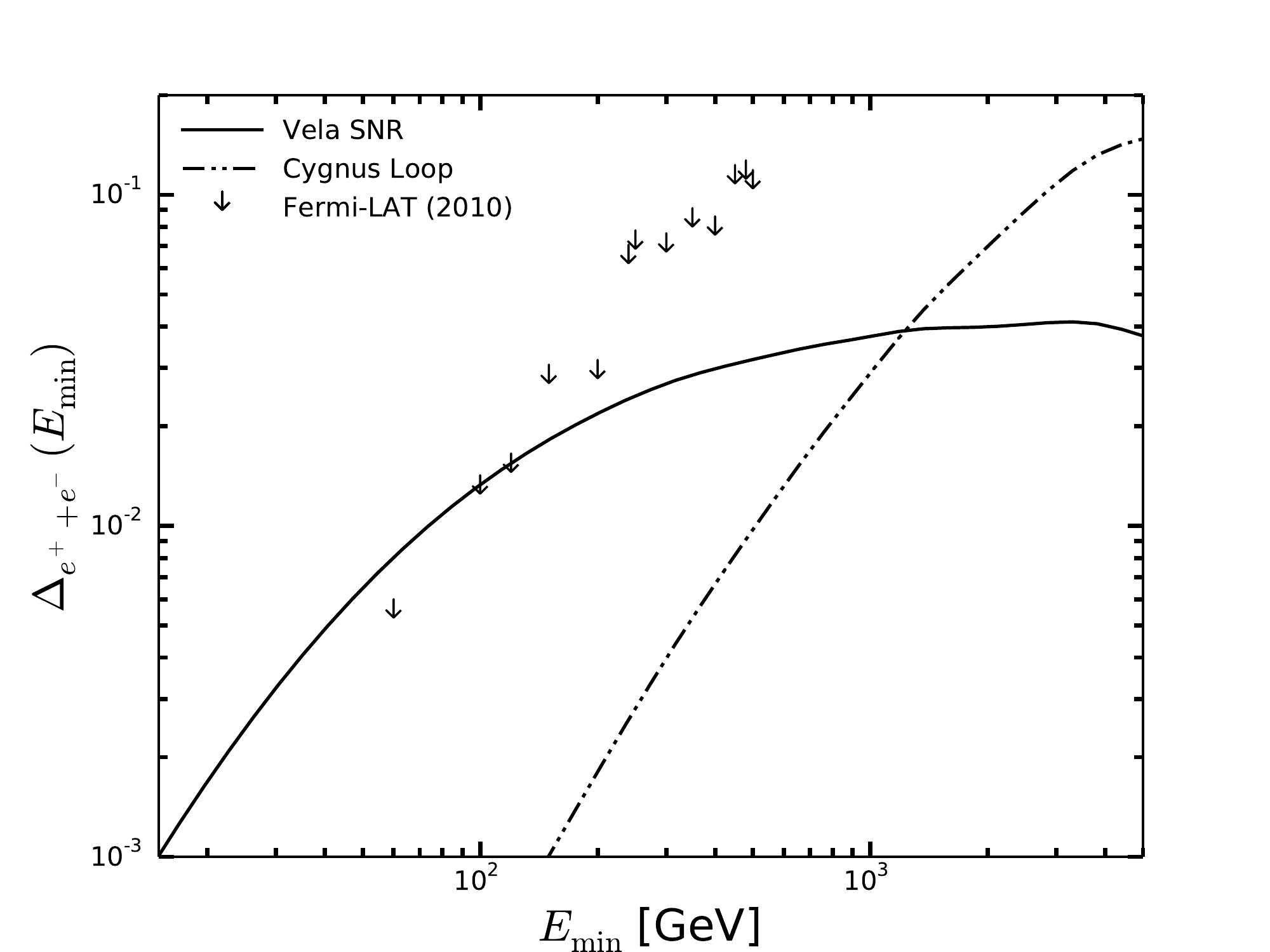}
\includegraphics[width=0.425\textwidth]{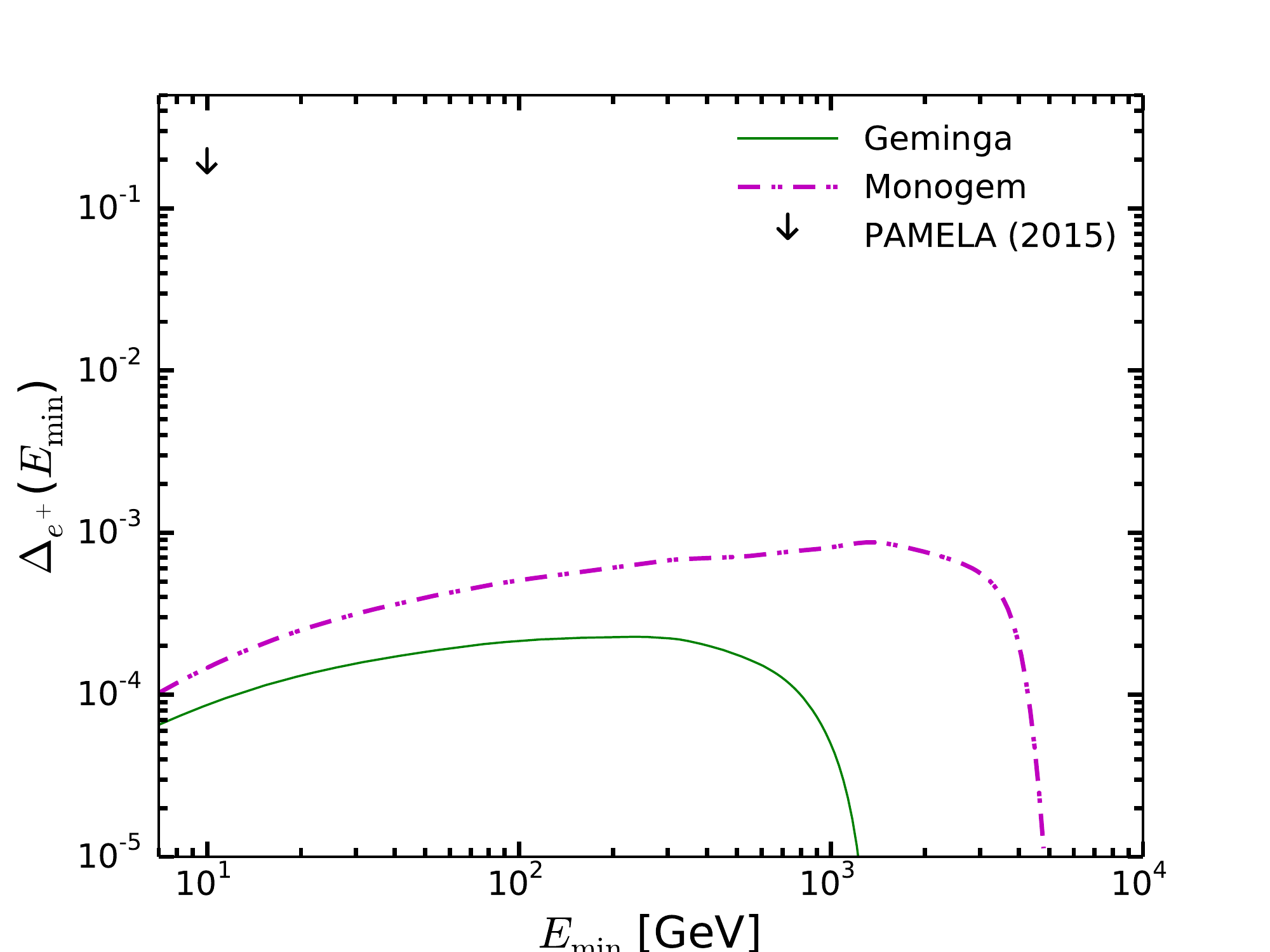}
\end{center}
\caption{\label{fig:dipole}Upper panel: predictions for the dipole anisotropies in the  $e^+ + e^-$ flux for the Vela SNR and the Cygnus Loop  as a function of $E_{\rm min}$ for the model with $R_{\rm cut}=0.7$~kpc. \textit{Fermi}-LAT upper limits \cite{2010PhRvD..82i2003A} are shown.
Lower panel: prediction for the dipole anisotropies in the $e^+$ flux for the Geminga and Monogem  PWNe  as a function of $E_{\rm min}$ for the model with $R_{\rm cut}=0$~kpc. PAMELA upper limit in \cite{Adriani:2015kfa} is shown.
}
\end{figure}
With the first model we aim to analyze the role of single near SNRs (in particular the Vela SNR) in the high energy flux and, consequently, in the electron plus positron anisotropy. Fluxes are fitted considering a secondary component, the contribution from single PWNe in the ATNF catalog, a smooth distribution of SNRs with  $R_{\rm cut}=0.7$~kpc, the contribution from single near SNRs in the Green catalog with $R\leq R_{\rm cut}$. Among the near SNRs, Vela is treated separately from the other sources as detailed in  ~\cite{new}. For this analysis, its spectral index is fixed to $\gamma=2.5$, its distance to 0.293 kpc and its age to 11.4 kyr. The results of our fit are presented in Fig.~\ref{fig:fluxes}. A number of free parameters is used to fit our model to the data. This includes a normalization for the secondary component, a common spectral index and efficiency $\eta$ for all the PWNe, a normalization for the \textit{near} component, the magnetic field for the Vela SNR, and a spectral index and a free normalization for the smooth SNR distribution. More details on the fit parameters are given in  ~\cite{new}. The fit to the AMS-02 fluxes is remarkably good, with a reduced $\chi^2/$d.o.f.$= 38/89$. The role of near SNR, in particular Vela, in shaping the $e^+ + e^-$ fluxes (left panel) is evident for $E\gtrsim 300$~GeV. We thus compute the corresponding dipole anisotropy for the Vela SNR and for the Cygnus Loop, which dominates the contribution of the near SNRs with $R<R_{\rm cut}=0.7$~kpc. 
In Fig.~\ref{fig:dipole} (upper panel) the integrated $e^+ + e^-$ anisotropy as a function of the $E_{\rm min}$ for Vela and Cygnus Loop of the model in Fig.~\ref{fig:fluxes} are shown. The arrows correspond to the \textit{Fermi}-LAT upper limits in  ~\cite{2010PhRvD..82i2003A}. The predicted anisotropies grow with $E_{\rm min}$ and reach the maximum value of $\Delta_{e^+ +e^-}= 0.04$ for the Vela SNR and $0.15$ for Cygnus Loop at TeV minimum energies. For $E_{\rm min}$ below about $150$~GeV the upper limits are at the same level of the prediction for the Vela anisotropy. Thus, present \textit{Fermi}-LAT upper limits start  to test some of the models (see also  ~\cite{new}) for the Vela SNR that are compatible with the flux data. Future results from the full statistics of \textit{Fermi}-LAT data, as well as ongoing experiments such as DAMPE and CALET \cite{2011NIMPA.630...55T}, will improve the potentiality for the anisotropy to explore and eventually exclude some of the models that explains the $e^\pm$ fluxes. 
To explore the role of the collection of all sources in this model, we show in Fig.~\ref{fig:countl} the interstellar intensity of the $e^+ + e^-$ flux as a function of the direction in the sky in Galactic coordinates for growing minimum energies. The result obtained with Eq.~\ref{eq:intensity} is shown by means of its percentage difference between the mean intensity from the entire source collection. The maximim of the intensity (yellow dot)  is found to be a direction very close to Vela (black dot) for $E_{min}=126$ and $E_{min}=661$~GeV (top panels). At higher energies, the interplay between the Vela, Cygnus Loop and the other sources shifts the maximal intensity in direction of Cygnus Loop (bottom panels). 
%
%

The second model aims to analyze the role of the most powerful PWNe among our collection of ATNF sources and, in particular, the resulting positron anisotropy. The difference with the previous model is that we consider a smooth distribution of SNRs all over the Galaxy, thus $R_{\rm cut}=0$~kpc, and none of the SNRs in the Green catalog. Therefore, the only single sources are the PWNe. For example Geminga and Monogem PWN, for which we present Fig.~\ref{fig:dipole} (lower panel) the integrated dipole anisotropies in the $e^+$ flux, togheter with the upper limit obtained with PAMELA data in  ~\cite{Adriani:2015kfa}.
The maximum anisotropy is given by the Monogem PWN at about $1$~TeV. Geminga gives a lower anisotropy due to its age ($343$~kyr vs. $111$~kyr for Monogem, see discussion in \cite{new}).   
The predicted anisotropy is more than three orders of magnitude below the $\Delta_{e^+}$ upper limit. The difference between predictions and upper limits is similar when computing the anisotropy in the positron to electron ratio to compare with AMS-02 upper limits \cite{PhysRevLett.110.141102}. This gap suggests that likely present or forthcoming data on positron anisotropy will not have the sensitivity to test the properties of ATNF PWNe that explains the AMS-02 data.

\bibliography{ECRS16_biblio}
%
%
%

\end{document}